\documentclass[fleqn]{article}
\usepackage[english]{babel}
\usepackage[utf8]{inputenc}
\usepackage[T1]{fontenc}

\usepackage{tikz}
\usepackage{forloop}
\usepackage{multicol}
\usepackage{multirow}

\usetikzlibrary{snakes}
\usepackage{pgflibraryshapes}

\usepackage{amsmath}
\usepackage{amssymb}
\usepackage{amscd}
\usepackage{amsthm}
\usepackage{cmll}
\usepackage{stmaryrd}
\usepackage{latexsym}
\usepackage{color}
\usepackage{amsfonts}
\usepackage{txfonts}
\newtheorem{proposition}{Proposition}

\newtheorem{definition}{Definition}

\usepackage{url}
\usepackage{hyperref}

\usepackage{array}

\usepackage[numbers]{natbib}

\usepackage{epsfig}  
\usepackage{setspace}
\usepackage{fancyhdr}
\usepackage{graphicx}

\usepackage{listings}

\title{On Simplifying Dependent Polyhedral Reductions}

\author{Sanjay Rajopadhye, Colorado State University}
\date{}
\begin{document}
\maketitle
\begin{abstract}
  \emph{Reductions} combine collections of input values with an associative
  (and usually also commutative) operator to produce either a single, or a
  collection of outputs.  They are ubiquitous in computing, especially with
  big data and deep learning.  When the \emph{same} input value contributes to
  multiple output values, there is a tremendous opportunity for reducing (pun
  intended) the computational effort.  This is called \emph{simplification}.
  \emph{Polyhedral reductions} are reductions where the input and output data
  collections are (dense) multidimensional arrays (i.e., \emph{tensors}),
  accessed with linear/affine functions of the indices.

  Gautam and Rajopadhye~\cite{sanjay-popl06} showed how polyhedral reductions
  could be simplified automatically (through compile time analysis) and
  optimally (the resulting program had minimum asymptotic complexity).  Yang,
  Atkinson and Carbin~\cite{yang2020simplifying} extended this to the case
  when (some) input values depend on (some) outputs.  Specifically, they
  showed how the optimal simplification problem could be formulated as a
  bilinear programming problem, and for the case when the reduction operator
  admits an inverse, they gave a heuristic solution that retained optimality.

  In this note, we show that simplification of dependent reductions can be
  formulated as a simple extension of the Gautam-Rajopadhye backtracking
  search algorithm.

\end{abstract}

\textbf{Keywords:} Polyhedral model, Reduction, Scheduling

\section{Background}

This note uses the notation from the previous work~\cite{sanjay-popl06,
  yang2020simplifying}, that we encourage the reader to review.  We briefly
summarize and clarify the core ideas in these papers.  Consider the following
two equations.

\begin{eqnarray}
  P[i] \label{eq:a}
  &=&
      \left\{
      \begin{array}{lcl} i=0&:& Q[i]\\
        i>0&:&\displaystyle \sum_{j=i}^{2i-1}Q[i] \end{array}\right. \\[2mm]
  X[i]\label{eq:x}
  &=&
      \left\{
      \begin{array}{lcl} i=0&:& f(i)\\[2mm]
        i>0&:&\displaystyle
               f\left(\sum_{j=0}^{i-1}X[i]\right) \end{array}\right.
\end{eqnarray}

These equations specify that each element of a one dimensional array
(respectively, $P$ and $X$) is obtained by \emph{reducing} (the operator is
addition) a subset of values of an input array (respectively, $Q$ and $X$).
The arrays have size $N$, viewed as a parameter of the program/equation, and
we seek to optimize the asymptotic execution time of our program as a function
of $N$.  The first equation defines an \emph{independent} reduction ($P$ is
output and $Q$ is input), while the second equation is a \emph{dependent}
reduction: $X$ is both input and output and appears on both the left and right
hand side.

Since the $i$-th element of the output involves the reduction of $i$ values,
the nominal complexity of each equation is $O(N^2)$.  We can recognize that
these summations are similar to prefix-sums: all the values (except the
last/first ones) contributing to the $i$-th output, also contribute to the
$(i-1)$-th output.  Simplification exploits this fact to compute each result
with a \emph{single} operation (in $O(1)$ time), thereby reducing the
asymptotic complexity to $O(N)$ as shown in the equations below.

\begin{eqnarray}
  P[i] \label{eq:a1}
  &=&
      \left\{
      \begin{array}{lcl} i=0&:& Q[i]\\
        i>0&:& A[i-1]+B[2i] -B[i] \end{array}\right. \\
  X[i]\label{eq:x1}
  &=&
      \left\{
      \begin{array}{lcl} i=0&:& f(i)\\
        i>0&:&f(X[i-1]) \end{array}\right.
\end{eqnarray}

\textbf{Polyhedra:} A polyhedron (polytope) $\cal D$ is the intersection of a
number of half-spaces or inequalities of the form $cz+\gamma\geq 0$ called
\emph{constraints}.  Some constraints may either be equalities (i.e., the
intersection of both $cz+\gamma\geq 0$ and $cz+\gamma\leq 0$), or ``thick
equalities,'' (the intersection of $cz+\gamma\geq 0$ and $cz+\gamma'\leq 0$,
for $\gamma'\neq\gamma$).  When $\cal D$ has such thick equalities, we say
that it saturates the constraint $\langle c, \gamma$, os simply saturates $c$.
Along any vector, $\rho$ such that $c\rho\neq 0$, $\cal D$ has only a bounded
number of points.  The intersection of all such saturating constraints is
denoted by $\cal L(D)$, and is the smallest linear subspace that contains
$\cal D$.

A \emph{facet}, $\cal F$ of a polyhedron $\cal D$ is its intersection with the
\emph{equality} $a z+\alpha = 0$ associated with exactly one constraint.  We
say that the facet \emph{saturates} the constraint $\langle a, \alpha\rangle$.
More than one constraint may be saturated, and this yields \emph{faces}.  The
concept of ``thickness'' can be extended to faces too, and Gautam and
Rajopadhye~\cite{sanjay-popl06} define the \emph{thick face lattice} of
$\cal D$ which is a critical data structure during simplification.
Zero-dimensional faces are called \emph{vertices}, 1-dimensional faces are
\emph{edges}, and $\cal D$ itself is the topmost face (it's children are the
facets).  Faces are arranged level by level, and each face saturates exactly
one constraint in addition to those saturated by its
``parent.''\footnote{Admittedly, the notion of a parent is ambiguous in a
  lattice, but since the faces will be visited recursively in a top down
  manner, the call tree of this recursion will provide the necessary context
  to identify the constraint uniquely.}

\textbf{Parameters, volume and complexity:} A polyhedron may have one or more
designated indices, like $N$ above, called its size
parameter(s).\footnote{Although not explicitly stated, Gautam and
  Rajopadhye~\cite{sanjay-popl06} assumed a single size parameter.}  There is
no upper bound on parameters, and they allow us to define an unbounded family
of polytopes, one for each value of the parameter.  The \emph{volume},
cardinality, or the number of integer points, in such a parametric polytope is
known to be a polynomial function of the parameter, and this polynomial is the
asymptotic complexity of a program that performs a contant time operation at
every point in the polyhedron.  The degree of this polynomial, also called the
number of (free) dimensions of the polyhedron, is the number of indices in
$\cal D$, less the number of linearly independent thick equalities.

\textbf{Equations, reductions, and their reuse and share space:} For the scope
of this note, we seek to simplify equations of the form
\begin{equation}\label{eq:red}
  \forall z\in {\cal D}: Y[B z] = \bigoplus e(z) ~~~=~~~ \bigoplus X[A z] 
\end{equation}
Here, $e$ is some expression, called the reduction body, and there is no loss
of generality in assuming that it simply reads an input array $X$.  $A$ and
$B$ are \emph{linear} access functions matching the number of dimensions of
$\cal D$, $X$ and $Y$, as appropriate.  Reductions combine multiple values to
produce multiple answers, and this is accomplished by a many-to-one (i.e.,
rank-deficient) linear \emph{write access}, $B$, called the \emph{projection}
of the reduction.  We say that the value of $e(z)$ \emph{contributes} to the
answer at $Y[B z]$.  The image of $\cal D$ by $B$, is the set of results
produced by the reduction, and is the \emph{domain} of $Y$, denoted by
${\cal D}_Y = B({\cal D})$.

Simplification is possible only if the \emph{same} input value is read at
multiple points in $\cal D$.  It is well known~\cite{fortes-moldovan,
  sanjay-fst-tcs, wong-delosme} that such \emph{reuse} occurs when $A$ is
rank-deficient: $z$ and $z'$ access the same value of $X$, iff $Az=Az'$, or
$z-z'$ is a linear combination of the \emph{basis vectors} of the null space
of $A$.  The \emph{reuse space} of our expression $e$, which we denote as
${\cal R}(e)$ is just the null-space of $A$, When an expression is to be
evaluated only at points in some domain $D$, its \emph{share space}
${\cal S} ({\cal D}, e)$, is defined to be ${\cal L(D)} \cap {\cal R}(e)$.
This is a linear space.\footnote{When it is more than one dimensional, there
  are infinitely many choices for the basis vectors.}

\section{Reduction Simplification}
\label{sec:gr06}

For clarity of explanation, we first assume that the reduction operator admits
an inverse, $\ominus$.  Later, we consider noninvertible operators.  We
simplify Eqn.~\ref{eq:red} recursively, going down the thick face lattice,
starting with $\cal D$, and at each step we simplify Eqn.~\ref{eq:red}, but
restricted to $\cal F$.

The key idea is that exploiting reuse along $\rho\in{\cal S}({\cal F}, e)$
avoids evaluating $e$ at most points in $\cal F$.  Specifically, let
${\cal F}'$ is the translation of $\cal F$ along $\rho$, and
${\cal F}'\backslash {\cal F}$ and ${\cal F}\backslash {\cal F}'$ be their
differences.  The union of these two is also the union of the thick facets of
$\cal F$.  Exploiting reuse along $\rho$ converts the original equation to a
set of \emph{residual computations} defined only on (a subset of) the facets
of $\cal F$.  All the computation in ${\cal F}\cap {\cal F}'$ is avoided.  To
understand the details, we first define two labels on faces (remember that in
this recursive traversal, every face $\cal F$ is associated with a unique
constraint, $\langle c, \gamma\rangle$).

First, a face ${\cal F}$, is said to be a \emph{boundary} face if its image by
$B$, $B({\cal F})$ is also a face of ${\cal D}_Y$, i.e., it contributes to a
``boundary'' of the result.  This happens if $Bc=0$.  Second, and with respect
to any reuse vector $\rho$, we define a face to be \emph{inward} if $c\rho>0$
(and respectively, \emph{outward} if $c\rho<0$ and \emph{invariant} if
$c\rho=0$).

A preprocessing step ensures that there are no invariant boundary faces.  We
do not perform any residual computations on the outward boundary, and on the
invariant facets of $\cal F$.  The other residual computations are used in
the following way.
\begin{itemize}\itemsep 0mm
\item The inward boundary faces are used to initialize the final answer.
\item The results of inward non-boundary faces are combined with
  $Y[B(z-\rho)]$ using the operator $\oplus$.
\item The results of outward non-boundary faces are combined with
  $Y[B(z-\rho)]$ using the operator $\ominus$.
\end{itemize}

\textbf{Optimality} At each step of the recursion, the asymptotic complexity
is reduced by exactly one polynomial degree, because facets of $\cal F$ have
one fewer free index.  Furthermore, at each step, the faces saturated by the
ancestors ensure that the new $\rho$ is linearly independent of the previously
chosen ones.  Hence, the method is optimal---the reduction in asymptotic
complexity is by a polynomial whose degree is the number of dimensions of the
feasible reuse space of the original domain, ${\cal L(D)} \cap {\cal R}(e)$,
and all available reuse is fully exploited.  This holds regardless of the
choice of $\rho$ at any level of the recursion (all roads lead to Rome) even
though there are infinitely many choices in general.

\paragraph{Handling operators without inverses and impact on optimality} Many
algorithms, particularly in dynamic programming, perform reductions with
operators like the $\min$ and the $\max$, which do not admit an inverse.  To
handle such equations, we must ensure that the residual computation whose
results are combined with $\ominus$ must have an empty domain, i.e., the
current face $\cal F$ soes not have any non-boundary outward facet, i.e.,
$c_i\rho\geq 0$ holds for all non-boundary facets.

There are two implications of this.  First, this means that the feasible space
of legal reuse vectors $\rho$ is no longer the linear subspace
$\in{\cal S}({\cal D}, e)$, but rather, must satisfy additional linear
inequalities.  Indeed, the feasible space may even be empty, and we may not be
able to exploit all available reuse.  For example, if the operator in
Eqn.~\ref{eq:a} is max, and we choose $\rho=[1,0]$, it makes the lower bound
$j\geq i$, an outward facet, while $\rho=[-1,0]$, makes the upper bound
$j\leq 2i$ outward.  Hence, this equation cannot be simplified and its
complexity will remain $O(N^2)$.

The second consequence is that, as the above algorithm recurses down the thick
face lattice, the choice of the $\rho$ at an earlier level may affect the
feasible space of lower levels, and hence the recursive algorithm is not
optimal.  Gautam and Rajopadhye~\cite{sanjay-popl06} solve this problem by
observing that the infinite feasible space can be partitioned into equivalence
classes based on the labels they assign to the non-boundary facets.  There are
because there are finitely many faces, and each has finitely many possible
labels, and hence a backtracking search over the thick face lattice,
formulated as a dynamic programming algorithm leads to an optimal choice of
$\rho$'s.

\section{Simplifying Multiple Statement Reductions}
\label{sec:yang20}

Recently, Yang, Atkinson and Carbin~\cite{yang2020simplifying} extended this
work to \emph{systems} of \emph{dependent} equations like the one in
Eqn.~\ref{eq:x}.  The main difficulty is that choosing a reuse vector $\rho$
to exploit introduces a new dependence in the program: $Y[Bz]$ depends on
$Y[B(z-\rho)$, which corresponds to a dependence vector $B\rho$.  This may
degrade the schedule of the program, or worse still, may even lead to a system
of equations that do not admit a schedule, e.g., had we chosen to exploit
reuse along $[-1,0]$ in Eqn.~\ref{eq:x}.

Yang et al.\ resolve this problem by combining the simplification and
scheduling problem, building off a long history of research on scheduling in
the polyhedral model.  They use the state of the art formulation by Pouchet et
al.~\cite{pouchet-etal-cgo07, pouchet-etal-pldi08, pouchet-etal-popl11} that
formulates multidimensional scheduling as a single linear optimization
problem.  For each variable/equation $X_i$ defined over a $d_i$-dimensional
domain, there is an $(1+d_i)\times (1+d_i$ matrix of schedule coefficients
 $\Theta_i$, and the causality constraints are translated to linear
inequalities defining a feasible space.  Many objective functions are used in
the literature, the most common being one that simultaneously optimizes for
locality and parallelism~\cite{uday-pldi08}.

However, there are two difficulties in doing this directly.  The first is
that, in the Gautam-Rajopadhye recursive algorithm, the $\rho$ vectors are
chosen one by one, and for each chosen introduces $B\rho$ as a new dependence
in the program that was not present in the original program.

Yang et al.\ resolve this by first formulating the Gautam-Rajopadhye algorithm
as a \emph{single} linear programming problem, by setting us simultaneous
constraints that must be satisfied for \emph{each face} of $\cal D$.  Next,
they modify the scheduling constraints so that the unknown $\rho$ vectors
appear in the causality constraints.  Because of this, the formulation does
not remain a linear programming problem, but becomes \emph{bilinear}.  They
couple this with a linear objective function that minimizes the asymptotic
complexity.

They concede that the solution of this problem may be difficult in the general
case, but propose a very simple heuristic that works really well for many of
the examples and algorithms they encounter in statistical machine learning.
They do not provide data about the execution time of their (non-heuristic)
implementation of the bilinear programming formulation.

\section{Simplifying dependent reductions as a backtracking search}
\label{sec:dep}
We now describe our main result.  We show how to solve the problem of
simplifying dependent reductions, like the ones tackled by Yang et al., by
extending the Gautam-Rajopadhye backtracking search algorithm.  It relies on
the early work on polyhedral scheduling~\cite{delosme-ipsen-europe} and builds
on a long history of scheduling~\cite{karp-etal, lamport-74, sanjay-fst-tcs,
  quinton-sanjay-tf, quinton-jvsp89, feautrier92a, feautrier92b}.

We first define \emph{compatibility} to capture the notion of the conditions
under which an additional dependence (e.g., one that is introduced by
simplification) does not introduce dependence cycles, and allows the program
to admit a legal schedule.
\begin{definition}
  Let $\cal T$, be the space of all legal schedules for a program.  We say
  that a new uniform self dependence vector $r$ on variable $Y$ is
  \textbf{compatible with} $\cal T$, or with the original program, if some
  feasible schedule $\Theta_Y$, respects the constraint $\Theta_Y r \succ 0$,
  where $\succ$ denotes the lexicographic order.  This holds iff
  \begin{equation}\nonumber
    \exists \Theta \in {\cal T} ~~ \mathrm{s.t.} ~~ \Theta r \succ 0        
  \end{equation}
  Otherwise, we say that $r$ \textbf{violates} $\cal T$.  More specifically, A
  reuse vector, $\rho$ for simplifying the reduction producing $Y$ is
  \emph{legal} if the uniform (self) dependence vector, $B\rho$ on the
  variable $Y$ is compatible with the original program.
\end{definition}

\begin{proposition}
  The feasible space of all multidimensional schedules for polyhedral pogram
  is a \emph{blunt, finitely generated, rational cone}. (see
  \url{https://en.wikipedia.org/wiki/Convex_cone}).
\end{proposition}
\begin{proof}
  A polyhedral set $\cal P$ is a cone iff for any point $x \in {\cal P}$, and
  for a positive scalars, $\alpha$, the point $\alpha x$, is also in
  ${\cal P}$.  The causality constraint states that for all pairs of iteration
  points, $z_X\in {\cal D}_X$ and $z_Y\in {\cal D}_Y$ such that $X[z_x]$
  depends on $Y[z_y]$, the time-stamp of the producer is strictly before the
  time stamp of the consumer.  Recall that a $d$-dimensional schedule $\Theta$
  satisfy causality iff the following constraint is satisfied (here
  $\delta = [0\ldots 0, 1]$ is a $d$-dimensional vector whose last element is
  1.
  \begin{equation}
    \label{eq:sched}  
    \Theta_X\left[\begin{array}{c} z_X\\p\\1] \end{array}\right] -
    \Theta_Y\left[\begin{array}{c} z_Y\\p\\1] \end{array}\right] \succ\delta
  \end{equation}
  Now if $\Theta$ is a legal schedule vector.\footnote{Note that we call it a
    \emph{vectors} even though it is a set of matrices, one for each variable
    in the program, since they constitute the unknown variables in the optimal
    scheduling linear program.} then it is easy to see that $\alpha \Theta$,
  for any positive $\alpha$ also satisfies~\ref{eq:sched}.  Indeed it is just
  a $\alpha$-fold slowdown of $\Theta$.  It is also easy to show that the cone
  is \emph{blunt}, i.e., it does not contain the origin, and because the
  schedule coefficients are integers, it is \emph{finitely generated}.
\end{proof}

We now formulate the additional legality conditions for reuse vectors used
during simplification.  Consider the cone defining the feasible schedule
$\cal T$ of a program (system of equations) prior to simplification.  Let its
projection on the dimensions representing the variable $Y$) be $\cal C$, and
let $\cal C$ have $m$ generators, $g_1$ \ldots $g_m$.

\begin{proposition}
  Simplifying the equation for $Y$ using a reuse vector, $\rho$ is legal iff,
  for some generator, $g_i$ of $C$, $g_i B\rho \geq 0$.
\end{proposition}

Thus, the legality conditions for reuse vectors used during the recursion in
the Gautam-Rajopadhye algorithm now become the disjunction of $m$ convex
constraints.  There are possibly $m$-fold more choices to explore (but with
the possibility of early termination), and once again, the optimality argument
carries over.

\bibliographystyle{plain}
\bibliography{main}

\end{document}